\documentclass[jkps,twoside,twocolumn,epsfig ]{revtex4}
\usepackage{graphicx}

\begin{document}
\title[] {Financial Networks in the Korean Stock Exchange Market }
%\Corresponding author{Tel.: +82-51-620-6354; fax: +82-51-611-6357}
\author{Seong-Min \surname{Yoon$^{}$}}
\affiliation{ $^{}$Division of Economics, Pukyong National University,\\
Pusan 608-737, Korea}
%\author{J. S. \surname{Choi$^{}$}}
%$\author{Myung-Kul \surname{Yum$^{c}$}}
\author{Kyungsik \surname{Kim$^{}$}}
\thanks{E-mail: kskim@pknu.ac.kr;\\
Tel: +82-51-620-6354; Fax: +82-51-611-6357}
\affiliation{
Department of Physics, Pukyong National University, \\
Pusan 608-737, Korea \\}
%$^{c}$Department of Pediatric Cardiology, Hanyang University,\\
%Kuri 471-701, Korea \\}

\received{28 February 2005 }

\begin{abstract}

We investigate the financial network in the Korean stock exchange
(KSE) market, using both numerical simulations and scaling
arguments. We estimate the cross-correlation on the stock price
exchanges of all companies listed on the the Korean stock exchange
market, where all companies are fully connected via weighted
links, by introducing a weighted random graph. The degree
distribution and the edge density are discussed numerically from
the market graph, and the statistical analysis for the degree
distribution of vertices is particularly
found to approximately follow the power law.\\
\hfill
%$^{*}$Corresponding author.Tel.:+82-51-620-6354;fax:+82-51-611-6357\\
%$E-mail$:kskim@pknu.ac.kr(K.Kim)

\end{abstract}

\maketitle

%\section { Introduction }

Recently, many interests have been concentrated on small-world and
scale-free network models [1] in the physical, biological, social
and technological networks. The small-world network, proposed by
Watts and Strogatz [2], has provided the connected properties
represented by single component graphs. The real-world models are
indeed characterized by the small world and clustering property,
e.g., social networks [3], the internet [4], document retrieval in
www[5], scientific cooperation relations [6], social networks of
acquaintances [7] and of collaborations [8]. The static and
dynamic behaviors extensively have studied on small-world
networks, and the prominent topics for small-world networks really
have a direct application in statistical mechanics and polymer
physics [9]. Until now, the financial market networks are
extensively used in various types of financial applications [10].

Furthermore, the self-organized behavior of individuals,
companies, capitalists, or nations has concentrated on forming
macroscopic patterns such as commodity prices, stock prices, and
exchange rates. There have existed many statistical quantities
that play a crucial role in describing the properties and patterns
of several options in financial markets. One of the most important
quantities among them is the auto-correlation between companies.
Until now, there have been many researches to estimate numerically
the correlation in price changes, using the random matrix theory
influenced on the collective behavior in financial markets
$[11-13]$.

Our purpose of this paper is to investigate the financial network
in the KSE. We find the cross-correlation on the stock price
exchanges of all companies listed on the Korean stock exchange
market of 2003, and the degree distribution and the edge density
are discussed from the market graph.

\begin{figure}[]
\includegraphics[angle=90,width=8.5cm]{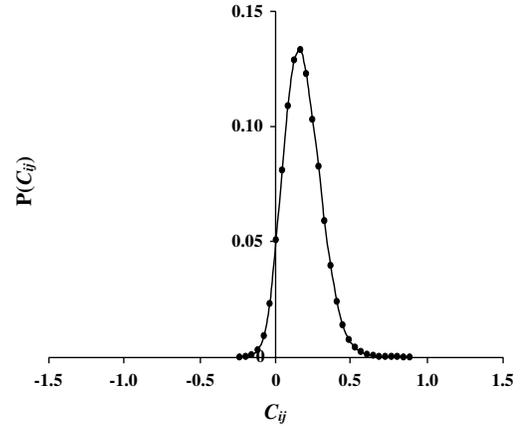}
\caption[0]{Plot of the distribution of correlation coefficients
in Korean stock exchange market.}
\end{figure}

%\section { Volatilities in the CTRW }

Let $R_i (t)$ be the return of stock price defined by
\begin{equation}
R_i (t ) = {\rm ln} [\frac{P_i (t + t_0 )} { P_i (t_0 )}],
\label{eq:a1}
\end{equation}
where $P_i (t)$ is the stock price of company $i$ at time $t$, and
$t_0$ is a time lag. Since the cross-correlation between
individual stocks are represented in terms of a matrix $C$, its
correlation coefficient is calculated as
\begin{equation}
C_{i,j} = \frac{<R_i R_j> - <R_i><R_j>}{\sqrt{(<{R_i}^2
>-<{R_j}>^2 )(<{R_i}^2 ) -<{R_i}>^2)}},
\label{eq:ba1}
\end{equation}
\begin{figure}[]
\includegraphics[angle=90,width=8.5cm]{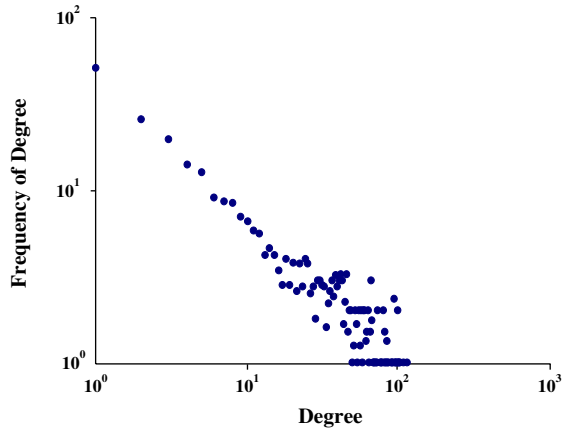}
\caption[0]{Plot of degree distribution of financial market
network for $\theta= 0.5$, where $P(\theta) \sim
{\theta}^{-\beta}$ with the scaling exponent $\beta=0.91$.}
\end{figure}
\begin{figure}[t]
\includegraphics[angle=90,width=8.5cm]{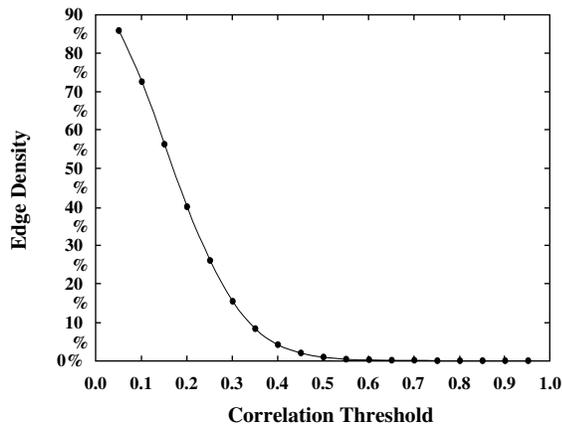}
\caption[0]{Plot of the edge density as a function of different
values of the correlation threshold $\theta$ in financial market
networks.}
\end{figure}
\begin{figure}[t]
\includegraphics[angle=90,width=8.5cm]{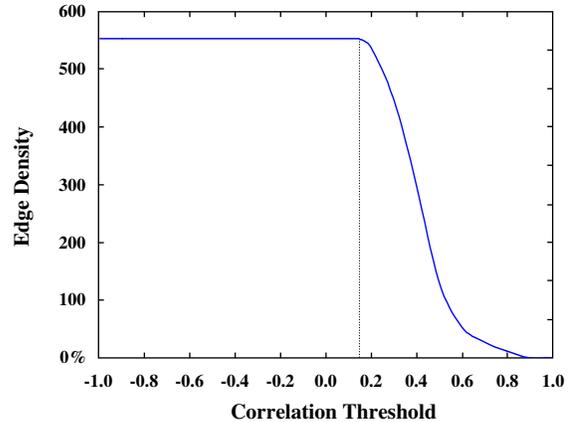}
\caption[0]{Plot of the largest connected edge density as a
function of the correlation threshold $\theta$ in financial market
networks. }
\end{figure}

\hfill\\
where the brackets denotes the time average over the transacted
period. From Eq. (1), the corresponding correlation coefficient
have one value between $[1,-1]$. If the coefficient
$C_{i,j}=1(-1)$, Two companies $i$ and $j$ are correlated
(anti-correlated) if the coefficient $C_{i,j}=1(-1)$, while they
are uncorrelated if $C_{i,j}=0$. It is well known the largest
eigenvalue is non-degenerate and real, since the matrix $C$ is
real and symmetric.

The market graph constitutes into the set $554$ of companies
traded in the the KSE. We analyze daily tick data for the period
Jan/$2003$-Dec/$2003$. First of all, as shown in Fig. $1$, we
found the distribution of correlation coefficients in Korean stock
exchange market, where $C_{i,j}$ vary in the range from $-1$ and
$1$.

In order to discuss the degree distribution and the edge density
in the market graph, we introduce as follows: The set of companies
represent the set of vertices of the graph. For any pair of
companies (vertices) $i$ and $j$, An connected edge for any pair
of companies (vertices) $i$ and $j$ is added to the market graph
if the corresponding correlation coefficient $C_{ij}$ is greater
than or equal to a specified threshold value $\theta$. The degree
probability that a vertex of the market graph has a edge $\theta$
follows a power law as
\begin{equation}
P(\theta)\propto {\theta}^{-\beta} .\label{eq:c3}
\end{equation}
For the degree distribution, we found the results of computational
simulation of our model with different threshold value. As the
correlation threshold value is increased larger than $\theta=0.4$,
the degree distribution resembles a power law. It is shown by
using least-squares method that the degree distributions of
financial market network for $\theta= 0.4$, $0.5$, and $0.6$ grow,
respectively, as a power law
%$P(\theta) \sim {\theta}^{-\beta}$
with scaling exponents $\beta=0.76$, $0.91$ (Fig. 2), and $1.15$.
We also found that the degree distribution in the range with
$\theta<0.4$ and $\theta>0.6$ do not follow a power law, different
to that of the recent work $[14]$. In Fig. $3$, The edge density
as a function of different values of the correlation threshold
$\theta$ in financial market networks is plotted in the range with
$\theta>0$. Fig. $4$ shows the largest connected edge density as a
function of the correlation threshold value $\theta$ in financial
market networks.

%\section { Conclusdions  }

In conclusion, we have studied the market graph in the Korean
stock exchange market and discussed the cross-correlation, the
degree distribution, and the edge density on the stock price
exchanges of all companies listed on the KSE of 2003. Especially,
We found that the degree distributions for $0.4\leq\theta\leq0.6$
follow a power law. It is in future expected that the detail
description of the market graph will be used to study the
extension of financial analysis in the Korean and foreign
financial markets.

\begin{acknowledgements}
This work was supported by Korea Research Foundation
Grant(KRF-2004-002-B00026).
\end{acknowledgements}

\end{document}